\documentclass[prb,twocolumn,10pt,twoside,superscriptaddress]{revtex4}
\usepackage[isolatin]{inputenc}
\usepackage{amssymb}
\usepackage{amsmath}
\usepackage[dvips]{graphicx}
\usepackage[dvips]{color}
\usepackage[tight,nice]{units}
\usepackage{xspace}

\newlength{\figwidth}

\newcommand{\Eq}[1]{Eq.~(\ref{eq:#1})}

\newcommand{\Eqs}[2]{Eqs.~(\ref{eq:#1}) and~(\ref{eq:#2})}
\newcommand{\Fig}[1]{Fig.~\ref{fig:#1}}
\newcommand{\Figs}[2]{Fig.~\ref{fig:#1} and Fig.~\ref{fig:#2}}
\newcommand{\Figure}[1]{Figure~\ref{fig:#1}}
\newcommand{\Sec}[1]{Section~\ref{sec:#1}}
\newcommand{\Secs}[2]{Sections~\ref{sec:#1} and~\ref{sec:#2}}
\newcommand{\PhiZ}{\ensuremath \Phi_{0}}
\newcommand{\U}{\ensuremath U}
\newcommand{\delw}{\Delta w}
\newcommand{\epsg}{\ensuremath \epsilon_\gamma}
\newcommand{\epsh}{\ensuremath \epsilon_h}
\newcommand{\gamb}{\ensuremath \gamma}
\newcommand{\gambbar}{\ensuremath \bar{\gamb}}
\newcommand{\gambtotal}{\ensuremath f}

\newcommand{\gamcn}[1]{\ensuremath \gamma_{C#1}}
\newcommand{\jc}{\ensuremath J_C}
\newcommand{\ja}{junction~\#2\xspace}
\newcommand{\jb}{junction~\#1\xspace}
\newcommand{\kb}{\ensuremath k_B}
\newcommand{\ket}[1]{|#1\rangle}
\newcommand{\hamdens}{\ensuremath{} {\cal H}}
\newcommand{\hamdensZ}{\ensuremath{} {\cal H}_0}
\newcommand{\htrans}{\ensuremath H_Z}
\newcommand{\htransZ}{\ensuremath H_{Z0}}
\newcommand{\htransZV}{\ensuremath \unit[0.011]{A/m}}
\newcommand{\lambdaJ}{\ensuremath \lambda_J}
\newcommand{\lmu}{\ensuremath l_{\mu}}
\newcommand{\mueff}{\ensuremath \bar{\mu}}
\newcommand{\omegaP}{\ensuremath \omega_P}
\newcommand{\thetaxy}{\ensuremath 35^\circ}
\newcommand{\thetayz}{\ensuremath 0.057^\circ}
\newcommand{\thrate}{\ensuremath \Gamma_\mathrm{th}}
\newcommand{\restmassprofile}{\ensuremath \tilde{\mu}}
\newcommand{\vortexmass}{\ensuremath{m_0}}
\newcommand{\vphitt}{\ensuremath \ddot{\varphi}}
\newcommand{\vphit}{\ensuremath \dot{\varphi}}
\newcommand{\vphixx}{\ensuremath \varphi''}
\newcommand{\vphix}{\ensuremath \varphi'}
\newcommand{\vphivx}{\ensuremath \partial \varphi_v / \partial x}
\newcommand{\vphivxfrac}{\ensuremath \frac{\partial \varphi_v}{\partial x}}
\newcommand{\vphi}{\ensuremath \varphi}
\newcommand{\w}{\ensuremath \omega_{0}}
\newcommand{\xz}{\ensuremath x_0}
\newcommand{\x}{\ensuremath x}
\newcommand{\xdepin}{\ensuremath x_{C2}}
\newcommand{\xdepinf}{\ensuremath x_{C1}}
\newcommand{\zetaV}{\ensuremath \unit[8]{m/A}}
\DeclareMathOperator{\asech}{arcsech}
\DeclareMathOperator{\sech}{sech}

\begin{document}

\setlength{\figwidth}{8.6cm}            

\title{Vortex qubit based on an annular Josephson junction containing
   a microshort}

\author{A.~N.~Price}
\affiliation{Physikalisches Institut III, Universit\"{a}t 
   Erlangen-N\"{u}rnberg, Erwin-Rommel-Str. 1, 91058 Erlangen, Germany}
\author{A.~Kemp}
\affiliation{Physikalisches Institut III, Universit\"{a}t 
   Erlangen-N\"{u}rnberg, Erwin-Rommel-Str. 1, 91058 Erlangen, Germany}
\author{D.~R.~Gulevich}
\affiliation{Department of Physics, Loughborough University,
   Loughborough, LE11 3TU, UK}
\author{F.~V.~Kusmartsev}
\affiliation{Department of Physics, Loughborough University,
   Loughborough, LE11 3TU, UK}
\author{A.~V.~Ustinov}
\affiliation{Physikalisches Institut, Universit\"{a}t Karlsruhe,
   Wolfgang-Gaede-Str. 1, 76131 Karlsruhe, Germany}

\begin{abstract}

We report theoretical and experimental work on the development of a vortex
qubit based on a microshort in an annular Josephson junction. The microshort 
creates a potential barrier for the vortex, which produces a double-well 
potential under the application of an in-plane magnetic field; The field 
strength tunes the barrier height. A one-dimensional model for this system is 
presented, from which we calculate the vortex depinning current and attempt 
frequency as well as the interwell coupling. Implementation of an effective 
microshort is achieved via a section of insulating barrier that is locally 
wider in the junction plane. Using a junction with this geometry we demonstrate
classical state preparation and readout. The vortex is prepared in a given 
potential well by sending a series of ``shaker'' bias current pulses through 
the junction. Readout is accomplished by measuring the vortex depinning 
current.

\end{abstract}


\maketitle


\section{Introduction} 

Superconducting qubits based on Josephson junctions are one of the most 
promising qubit
architectures in terms of scalability and ready integration
with semiconductor electronics.
Josephson junctions have been successfully utilized to build 
various types of qubits such as charge,\cite{nakamura99} 
phase,\cite{martinis02} and flux\cite{chiorescu03} qubits. 
The operation of these systems is based on quantum 
coherence of the charge state, the Josephson phase difference, or the 
magnetic flux state, respectively. Elements characteristic of
the charge and flux qubits are combined in a hybrid
qubit called quantronium.\cite{vion02} 

By contrast with other types of superconducting qubits, a vortex qubit 
is designed to exploit the coherent superposition of two
spatially separated states for a Josephson vortex within a long Josephson 
junction.\cite{wallraff00a,kemp02b,fistul03} These states correspond to
the minima of a double-well potential.
The landscape of potential energy experienced by the vortex along
the length of the junction can be constructed as desired by spatially
varying one or more of the following parameters: the junction
barrier thickness and hence the critical current 
density,\cite{shaju,shaju05,wallraff00phd,kim06,mclaughlin78, kato96} the 
magnitude of the in-plane
magnetic field,\cite{fistul03,kaplunenko04,carapella04} 
the curvature of the junction
centerline,\cite{wallraff00a, kemp02b, shaju06} and the width of the 
junction.\cite{gold01,benabdallah96,kemp06phd} 

\begin{figure}[htbp]
\resizebox{\figwidth}{!}{\includegraphics{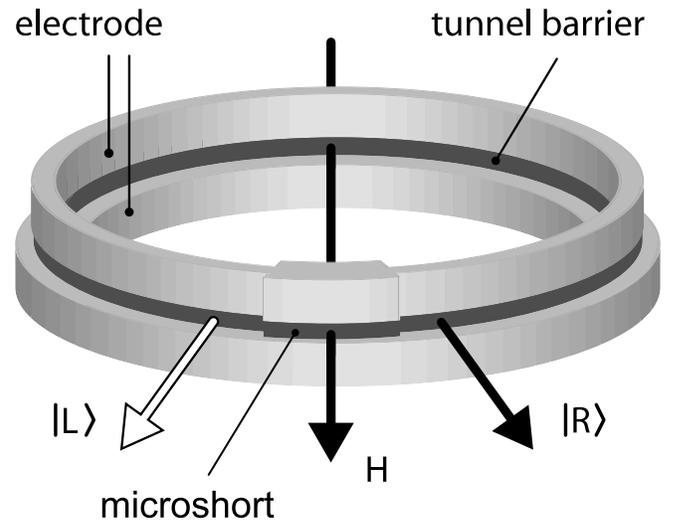}}
\caption{A vortex qubit consisting of a one-dimensional annular junction with
a microshort formed by a section of tunnel barrier of slightly greater 
in-plane width. Potential wells develop to the left~$\ket{L}$ and 
right~$\ket{R}$ of the microshort when a static magnetic
field~$H$, represented by the long arrow, is applied. A vortex at rest
at either minimum is in a classically stable state. The magnetic moment
of the vortex is indicated by a short arrow.
(For clarity, the schematic is not drawn to scale.)
}
\label{fig:qubit}
\end{figure}

In this article, we report theoretical considerations and 
experimental results concerning a suggested vortex qubit\cite{shaju} which 
consists of a long one-dimensional annular junction\cite{davidson85} 
containing a microshort.\cite{shnirman97b} For this type of qubit, the first 
quantitative fit between an analytical model and experimental data in the 
classical regime is presented. The investigated system is 
schematized in \Fig{qubit}. In a microshort qubit, competition between 
repulsion at the microshort and pinning by an in-plane magnetic 
field\cite{wallraff03,wallraff00phd} creates 
a double-well potential for the vortex\cite{shaju,shaju05}. In having a 
double-well potential, the microshort qubit resembles the flux 
qubit.\cite{friedman00,wal00} An advantage of qubit basis states
being localized in separate wells is that intrawell energy relaxation
does not cause a false readout result. The quantum state of the microshort
qubit is manipulated by applying a short pulse of magnetic field, which lowers
the height of the microshort-induced potential barrier and thus enables 
coherent oscillation between the basis states. This field plays the same 
role in controlling the tunneling amplitude as the external magnetic flux
in the early charge qubit.\cite{makhlin99}

We have enhanced a previously outlined design\cite{shaju} of the microshort 
qubit by implementing a microshort compatible with standard lithographic 
fabrication processes.\cite{kemp06phd} At a ``lithographic'' microshort, the 
in-plane width of 
the tunnel-barrier is fractionally larger than it is elsewhere in the junction, 
resulting in a locally enhanced critical current density per unit length. 
This planar structure aids the monolithic integration of microshort qubits in 
Rapid Single Flux Quantum (RSFQ) circuits, as does the junction being large 
enough for lithographic patterning and having a compatible critical current
density~$\jc$. With RSFQ logic as the interface between vortex qubits and 
room temperature electronics, the circuits could readily be scaled up to the 
large numbers needed for useful computations. We have previously 
reported\cite{kemp06phd} that the magnitude of the bias current required
to drive a vortex past a lithographic microshort was an order of 
magnitude larger than expected. In this article, we explain that the
discrepancy is due to interaction between the lithographic microshort and
magnetic field oriented transverse to the junction plane.

A number of characteristics of the proposed microshort qubit are tunable
during experiment. The applied in-plane magnetic field controls the 
height of the potential barrier, the separation of the potential minima,
and the frequency of coherent oscillation of the vortex. Also the
coupling between microshort qubits is adjustable. Josephson vortex qubits 
placed in a superconducting transmission line couple via virtual 
electromagnetic waves excited and absorbed by vortices.\cite{fistul03} This 
indirect interaction depends on the tunneling amplitude of each qubit.

Since quantum tunneling of a single vortex out of a metastable potential
well has been demonstrated,\cite{wallraff03} the next major step in
realizing a vortex qubit is observing coherent 
oscillation of a vortex in a double-well potential. Using the experimentally 
investigated system detailed in this article, we have observed for the 
first time a single vortex escaping from a metastable state by tunneling 
through a microshort-induced barrier.\cite{price10} However we limit the 
experimental data presented here to the classical regime, focussing on
state preparation and readout.

The theoretical section of this article,~\Sec{theory}, begins with a
derivation of the one-dimensional vortex potential for the proposed
qubit. An analytical expression is obtained for the depinning current
of a vortex over a microshort-induced potential barrier as a function of
bias current and in-plane magnetic field strength. Also, the attempt frequency
of the vortex in the presence of large magnetic field is derived for the cases 
of zero bias current and bias just below the critical current. From
the attempt frequency and the height of the potential barrier, we find the 
field dependence of the coupling between degenerate potential minima.

Experimental results obtained in the classical regime are presented
in \Secs{results}{disc}. We report measurements of the
vortex depinning current as a function of magnetic field strength that
indicate the presence of bistable vortex states. Our
readout scheme, based on the unique depinning current of each
state, is described along with results which confirm that the vortex
was prepared in a chosen initial state by means of shaker bias current
pulses. The final section,~\Sec{disc}, discusses 
how the vortex depinning current over a lithographic-microshort-induced 
potential barrier can be enhanced by flux trapped around both
junction electrodes or magnetic field applied transverse to the
junction plane. 

\section{Vortex in an annular microshort junction - Theory}
\label{sec:theory}

\subsection{One-dimensional model}

\begin{figure}[htbp]
\resizebox{\figwidth}{!}{\includegraphics{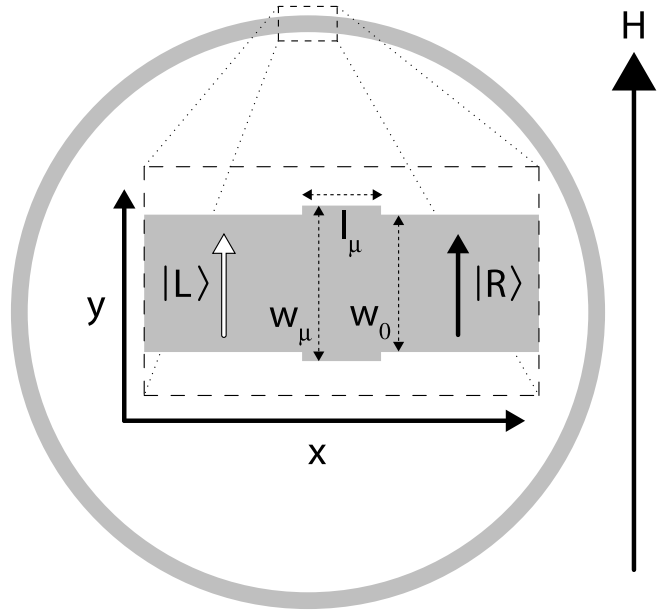}}
\caption{
Plan view of the tunnel-barrier layer of an annular junction with a 
lithographic microshort. The enlargement shows
the effective microshort in more detail, where
the barrier layer is fractionally wider than it is everywhere else 
in the junction. 
}
\label{fig:lith}
\end{figure} 

A homogeneous long Josephson junction is well described by the
one-dimensional sine-Gordon equation for the phase difference
$\vphi$ between the order parameters of the superconducting
electrodes:\cite{Gronbech91b,gronbech91c}
\begin{equation}
\label{eq:sge}
\vphixx - \vphitt - \sin\vphi = \gambtotal\,. 
\end{equation}
The spatial coordinate $x$ (see~\Fig{lith}) is in units of
the characteristic length scale,
the Josephson length~$\lambdaJ$, while time $t$ is normalized to
the inverse of the Josephson plasma frequency $\omegaP$. The perturbation 
\begin{equation}
\label{eq:pert}
\gambtotal(x)=\gamb + \frac{\partial{h_R}}{\partial{x}}
\end{equation}
incorporates the bias current~$\gamb$
and the current induced by the radial component~$h_R$
of the normalized external magnetic field $h=\kappa H/H_0$.
The bias current~$\gamb$ is normalized to the 
product of the critical current density~$\jc$ and the junction area. 
The characteristic field is given by
$H_0=\PhiZ/2\pi\mu_0d\lambdaJ$ where~$\PhiZ$ is the magnetic flux
quantum, $\mu_0$~is the vacuum permeability, $d$~is the magnetic
thickness of the junction, and~$\kappa$ is the geometric coupling
factor between the magnetic field applied and that in the junction. 
For a linear junction $\partial{h_R}/\partial{x}=0$,
but in the annular case $\partial{h_R}/\partial{x}=-(h/r)\sin(x/r)$
where~$r$ is the junction
radius.  Dissipative terms have not been included 
in~\Eq{sge} because, for the purposes of the experiments described in
\Secs{results}{disc}, the damping in our 
system at low temperatures is small enough to neglect.

In the absence of external field and applied bias current, one of
the solutions of~\Eq{sge} is that of the stationary Josephson
vortex:
\begin{equation}
\label{eq:vortexprofile}
\vphi^v=4 \arctan [\exp(\x-\xz)]\,.
\end{equation} 
The Josephson vortex is a topological soliton, behaving like a 
particle of normalized rest mass energy~$\vortexmass=8$ and center of
mass position~$\xz$. Here the unit of energy 
${\cal E}_0=\jc w_0 \lambdaJ \PhiZ/2\pi$ is the Josephson 
coupling energy of a small junction of
area~$w_0 \lambdaJ$, $w_0$ being the junction width.

To find the potential seen by a vortex in an annular junction,
we begin with the Hamiltonian corresponding to~\Eq{sge}:
\begin{equation}
\label{eq:sgeham}
\hamdensZ=\int_0^{l} \Bigl( \frac12 \vphix^2 + \frac12 \vphit^2 
   + 1- \cos\vphi + \gambtotal \vphi \Bigr) \,dx\,.
\end{equation}
Approximating the spatial phase profile~$\vphi(x)$ in~\Eq{sgeham} 
with~\Eq{vortexprofile}, that of a vortex, and carrying out the
integration over the entire length~$l=2\pi r$ of the junction, one
obtains the washboard potential:\cite{Gronbech91b}
\begin{equation}
\label{eq:annpot}
U(x_0) = - 2 \pi \Bigl( \gamb x_0 + h\sech\frac{\pi}{2r}\cos\frac{x_0}{r} 
             \Bigr)\,.
\end{equation}
The first term $U^\gamb=- 2 \pi \gamb x_0$ reflects the driving action
of the bias current on the vortex, while the second,
$U^h=- 2 \pi h\sech(\pi/2r)\cos(x_0/r)$, originates in the 
convolution of the vortex spatial magnetic field profile $\vphivx = 2 \sech x$ 
with the applied field component~$h_R$:
\begin{equation}
\label{eq:fconv}
U^h = - h_R * \vphivxfrac \,.
\end{equation}
Next we extend \Eq{annpot} 
by adding the contribution due to the lithographic microshort
depicted in~\Fig{lith}.

The characteristic energy scale~${\cal E}_0$ and hence the
Josephson vortex rest mass energy~$\vortexmass$ are proportional to 
the width of the junction. Thus one expects that a short
length~$\lmu$ of broader junction will act as a potential
barrier to the vortex, just as a microshort formed by decreasing the
tunnel barrier thickness would.
In the case where the lithographic microshort
length is smaller than the characteristic vortex size, $\lmu < \lambdaJ$,
it is important to consider the spatial distribution
of the vortex mass.

The spatial rest mass profile is found from the sine-Gordon 
Hamiltonian for a static vortex: \Eq{sgeham} with $\gambtotal=0$. 
The time-independent phase distribution,~\Eq{vortexprofile}, 
gives rise to equal magnetic field and Josephson
coupling energy densities, $\vphix^2/2$ and~$1-\cos\vphi$
respectively.
Hence the vortex rest mass
energy can be written as 
\begin{equation}
\label{eq:annmass}
\vortexmass=\int_0^{l} \restmassprofile \,dx
\end{equation}
where the spatial distribution of the vortex rest mass is
\begin{equation}
\label{eq:massprofile}
\restmassprofile = 4 \sech^2(x-x_0)\,.
\end{equation}

To calculate the change of vortex rest mass energy caused by altering
the junction width, we start with an appropriate Hamiltonian
\begin{equation}
\label{eq:varwham}
\hamdens=\int_0^{l} \frac{w(x)}{w_0} \Bigl( \frac12 \vphix^2 
   + 1- \cos\vphi \Bigr) \,dx\, ,
\end{equation}
neglecting the perturbations due to bias current and external 
magnetic field. Note that the energy normalization~${\cal E}_0$
remains constant since the factor $w(x)/w_0$ describing the
spatial variation of the junction width appears 
explicitly in~\Eq{varwham}. We only consider small width changes, 
so the phase distribution
$\vphi(x,y)=\vphi(x)$ remains radially independent.\cite{gold01}
One now sees that the vortex rest mass energy~$m_0$ in a long junction
of variable width
consists of the convolution of the spatial rest mass profile
belonging to a uniform junction with the local potential
energy per unit length of the variable width junction:
\begin{equation}
\label{eq:restmass}
\vortexmass=\restmassprofile*\frac{w}{w_0} \,.
\end{equation}

\begin{figure}[htbp]
\resizebox{\figwidth}{!}{\includegraphics{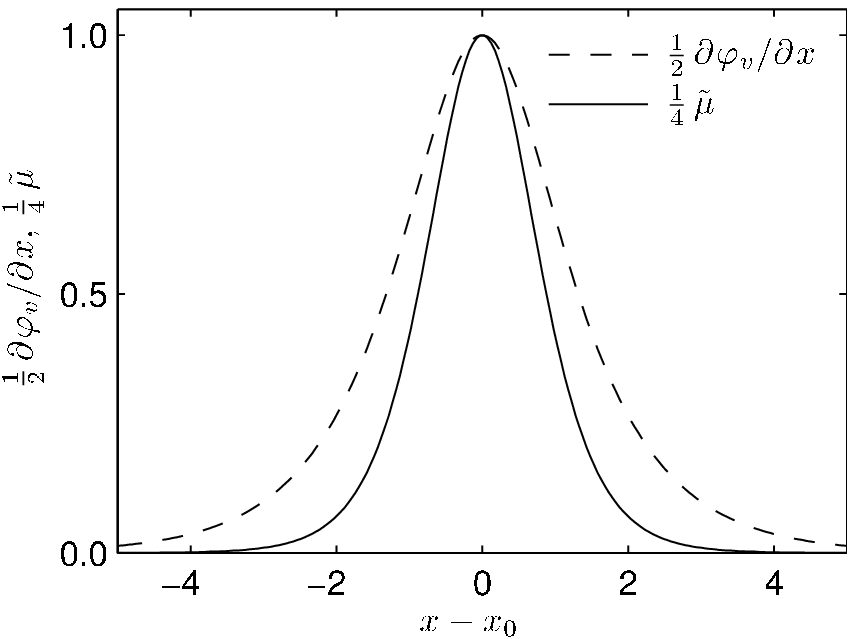}}
\caption{Comparison of the kernels~$\vphivx$ and~$\restmassprofile$
of the convolution 
integrals which respectively give rise to the magnetic field~$U^h(x_0)$ and 
microshort~$U^\mu(x_0)$ contributions to the vortex
potential. The rest mass profile 
$\restmassprofile\sim\sech^2(x-x_0)$ (solid line) 
of the vortex is spatially more tightly
confined than its magnetic field profile 
$\vphivx\sim\sech\,(x-x_0)$ (dashed line); Therefore the resultant
microshort barrier tends to be narrower, and the
associated
vortex quantum tunneling rate should be enhanced.\cite{kato96} }
\label{fig:sechsechsqprof}
\end{figure}

The kernel $\restmassprofile\sim\sech^2(x-x_0)$ of this convolution integral 
has a different shape to the magnetic field profile $\vphivx\sim\sech(x-x_0)$
in~\Eq{fconv}. The rest mass and magnetic field profiles are plotted together
in~\Fig{sechsechsqprof}; For the same amplitude, the rest mass profile 
is narrower. This means that the potential barrier caused by a microshort 
tends to be more spatially confined than a barrier produced by an external 
field. Consequently, quantum tunneling of a Josephson vortex through a 
microshort-induced barrier is expected to be enhanced.\cite{kato96}

Turning to the specific long junction geometry depicted
in~\Fig{lith}, the width variation is
\begin{equation}
\label{eq:rect_width}
\frac{w(x)}{w_0}=1+\frac{\delw}{w_0} \Pi(x/\lmu)
\end{equation}
where $\Pi(x)$ is the unit rectangle function and
$\delw=w_\mu-w_0$ is the amount by which the junction
width at the lithographic microshort,~$w_\mu$, is larger than the junction
width elsewhere. For short microshorts, $\lmu \ll \lambdaJ$, an equivalent 
representation using the Dirac delta function is
\begin{equation}
\label{eq:delta_width}
\frac{w(x)}{w_0}=1+\frac{\delw}{w_0} \lmu \delta(x) \, .
\end{equation}
Then the potential energy corresponding to the change in vortex
rest mass energy evaluates to
\begin{equation}
\label{eq:Umicro}
\U^\mu(x_0)=\mu \sech^2 x_0
\end{equation}
where $\mu=4\lmu\delw/w_0$ measures the strength of the microshort.
Combining~\Eq{Umicro} and~\Eq{annpot} gives the vortex potential
for an annular junction containing a microshort:
\begin{equation}
\label{eq:micropot}
U(x_0) = \mu \sech^2 x_0- 2 \pi \Bigl(\gamb x_0 
         + h\sech\frac{\pi}{2r}\cos\frac{x_0}{r}\Bigr) \, .
\end{equation}

\subsection{Double well potential}

\begin{figure}[htbp]
\resizebox{\figwidth}{!}{\includegraphics{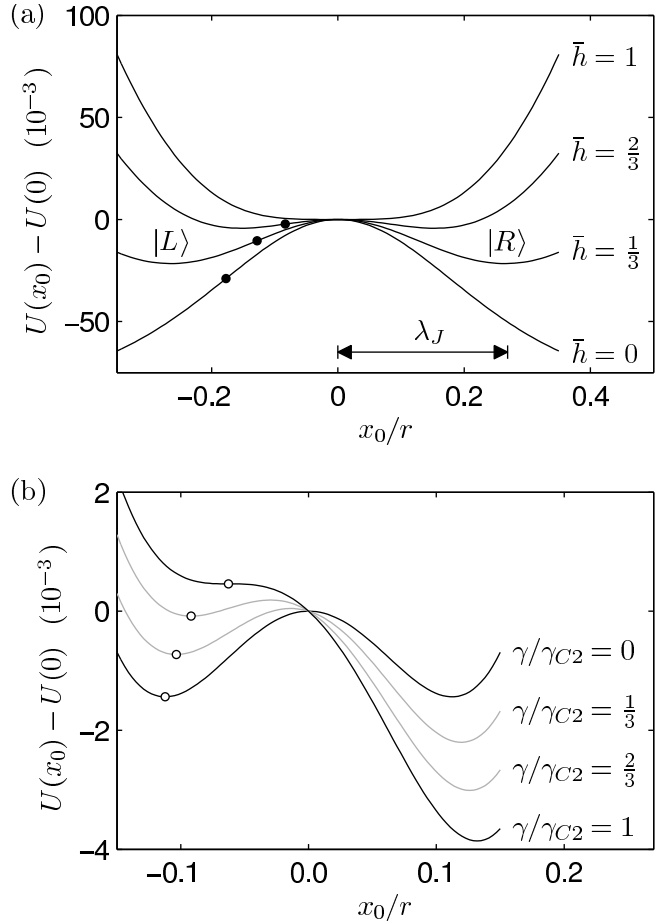}}
\caption{The vortex potential~$U(x_0)$ given by~\Eq{micropot}
for a long annular junction of radius $r=3.7$ containing a microshort of 
   strength $\mu=\unit[0.09]{}$. (a) The potential is plotted for four magnetic
   fields in the range $0 \leqslant \bar{h} \leqslant 1$ at zero bias
   current. The double-well potential is tunable: As the magnetic 
   field strength increases, both the
   height of the microshort-generated potential barrier and the
   physical separation between the stable vortex states~$\ket{L}$ 
   and~$\ket{R}$ decrease. A black circle marks the location~$\xdepin$
   at which an inflection point forms under the critical bias~$\gamcn{2}$.
   (b) The potential at $\bar{h}=0.8$ is shown for four bias currents in 
   the range $0 \leqslant \gamb \leqslant \gamcn{2}$. A white circle
   indicates the location of the left
   minimum, which becomes an inflection point at $\gamb=\gamcn{2}$.}
\label{fig:dwell}
\end{figure}

The vortex potential in the absence of bias current, \Eq{micropot} with 
$\gamb=0$, has two wells for magnetic fields in the range $0 < \bar{h} < 1$ 
where
\begin{equation}
 \bar{h} \equiv \frac{h \pi \sech(\pi/2r)}{\mu r^2} \, .
\end{equation}
\Figure{dwell}a contains plots of this potential for various values of
the magnetic field. The parameters~$\mu$,~$\lambdaJ$, and~$r$
were chosen to reflect the experimentally investigated system, which is 
detailed in~\Sec{setup}.
The height of the potential barrier shrinks with increasing
magnetic field strength. In this way, the barrier height can be
quickly controlled during experiment by, for example, passing current
through an appropriately oriented microstrip underneath
the junction. The field at which the barrier disappears
completely is found by calculating when 
the curvature~$U''(x_0)$ of the potential becomes 
positive at~$x_0=0$. The left~$\ket{L}$ and right~$\ket{R}$ wells 
of the potential constitute stable classical states for the vortex. 

The states are distinguished
by their critical, or depinning, current. This is the amount of bias current 
needed for the vortex to overcome the pinning potential. We denote the
depinning currents for the field- and microshort-induced barriers as
$\gamcn{1}$ and $\gamcn{2}$, respectively. As shown in
\Fig{dwell}b for field $\bar{h}=0.8$, a larger bias current tilts
the potential further. The left potential minimum, indicated by a white
circle, turns into a horizontal inflection point at the critical 
bias $\gamcn{2}$. The location at which the inflection point 
develops, $\xdepin$, is represented in \Fig{dwell}a by black circles. 
The depinned vortex gains kinetic energy as it moves through the microshort
and beyond in the direction of increasing~$x$, and the junction voltage
becomes nonzero. The vortex is metastable at bias currents slightly below
that required to transform its potential well into an inflection
point. This is due to thermal fluctuations and quantum tunneling, which
reduce the observable critical current. At a large enough magnetic 
field~$\bar{h}$, a vortex which has depinned from the left~$\ket{L}$ well 
will be retrapped by the field-induced potential barrier 
before the switch to the resistive state can be observed. This can be avoided 
during readout of the qubit state by decreasing the field before the 
depinning current is measured.

We now consider the critical current~$\gamcn{1}(h)$ 
associated with a vortex escaping over the potential barrier induced by the 
magnetic field. Increasing positive bias current moves the right 
minimum~$\ket{R}$ and the vortex farther in the positive $x$~direction.
At the critical bias, the vortex reaches the location~$\xdepinf = \pi r/2$,
where its magnetic moment is perpendicular to the applied field and the 
potential has a saddle point. Here the vortex depins and the bias
accelerates it.
In the case of a long annular junction with a weak microshort strength 
$\mu \ll 1$, it is 
reasonable to neglect the influence of the microshort on
this critical current.
Therefore\cite{wallraff00phd}
\begin{equation}
\label{eq:IcH1}
\gamcn{1}(h)\simeq \frac{|h|}{r} \sech \frac{\pi}{2r} \, .
\end{equation}

Next we derive the critical current~$\gamcn{2}(h)$
due to the microshort. In this case, we assume $x_0\sim 0$ 
and replace the cosine term in~\Eq{micropot} with its Maclaurin series up 
to $O(x_0^2)$, neglecting an additive constant: 
\begin{equation} 
\label{eq:Ux0}
U(x_0)\simeq \mu \left( \sech^2 x_0+ \bar{h} x_0^2\right) 
                    - 2 \pi \gamb x_0\,.
\end{equation}
At the critical bias~$\gamcn{2}$, the vortex depins at the location~$\xdepin$ 
where the potential has a saddle point:
\begin{align} 
U'(\xdepin)&=0 \quad \text{and} \label{eq:cond1} \\ 
\quad U''(\xdepin)&=0\,. \label{eq:cond2}
\end{align}
The condition given by~\Eq{cond2} leads to a quadratic equation in 
$\sech^2 \xdepin$ with the solution
\begin{align}
\sech^2 \xdepin &= \frac{1+\sqrt{1+3\bar{h}}}{3} \equiv \tau_1(\bar{h}) \,,\\
 \xdepin &= - \asech\, \sqrt{\tau_1} \,. \label{eq:xc}
\end{align}
The minus sign in~\Eq{xc} reflects the case of a vortex which moves in the 
positive~$x$ direction as it escapes over the microshort-induced
potential barrier. A larger applied field in the 
range~$0<\bar{h}<1$ results in a depinning location~$\xdepin$ which is
nearer to the microshort center at~$\x=0$.
After substituting~\Eq{xc} into~\Eq{cond1}, one 
obtains the field-dependent critical current for a vortex trapped by a 
microshort:
\begin{equation}
\label{eq:IcH2}
\gamcn{2}(h) = \frac{\mu}{\pi} \left(\tau_1 \sqrt{1-\tau_1} 
             - \bar{h} \asech \sqrt{\tau_1} \right) \, .
\end{equation}

\subsection{Attempt frequency}

A vortex in a potential well oscillates about its average position
with frequency~$\w$, known as the small amplitude oscillation frequency.
This is the frequency with which the vortex attempts to escape from
the well; In a junction with moderate damping it is proportional to 
the rate of thermal escape. [See \Eq{thesc}.] In the quantum
regime, the energy $\hbar\w$ separates the first excited state
of the vortex from its ground state within a single well. For a 
symmetric double-well potential, the frequency of coherent oscillation
of the vortex between the $\ket{L}$ and $\ket{R}$ states, $\Delta_0$, 
depends exponentially on the small oscillation frequency~$\w$ [as seen
from \Eqs{coupling}{action}]. In this section, we derive analytical
expressions for the attempt frequency~$\w$ at large fields 
$\bar{h} \lesssim 1$.

The frequency of attempted vortex escape varies as the square
root of the curvature at the potential minimum:
\begin{equation}
\label{eq:w}
\w=\sqrt{\frac{U''(x_1)}{\vortexmass}}\,.
\end{equation}
The location of the potential minimum is found by solving
\begin{equation}
U'(x_i)=0
\end{equation}
using the approximation $x \simeq \tanh x + \tanh^3 x / 3$, which is valid
for small~$x_i$ and therefore large normalized fields~$\bar{h}$. The
approximation leads to an analytically solvable cubic equation in $\tanh x_i$:
\begin{equation}
\label{eq:cubic}
\tanh^3 x_i + \tau_2 \tanh x_i = \tau_3 
\end{equation}
where, for clarity, we define
\begin{equation}
\tau_2(\bar{h}) \equiv -\frac{1-\bar{h}}{1+\bar{h}/3} \quad \text{and} \quad
\tau_3(\bar{h}) \equiv \frac{\pi \gamb}{\mu(1+\bar{h}/3)}\,.
\end{equation}
The discriminant of the cubic equation,
$D=(\tau_2/3)^3 + (\tau_3/2)^2 $, 
equals zero at the critical 
bias~$\gamcn{2}$, where all roots are real and two are equal. From this, one
obtains an approximation to~\Eq{IcH2},
\begin{equation}
\label{eq:IcH2simple}
\gamcn{2}(\bar{h}) 
   \simeq \frac{2\mu(1-\bar{h})^{3/2}}{3\sqrt{3}\pi(1+\bar{h})^{1/2}}
\,,
\end{equation}
whose relative error is under 0.7\% over the range of 
magnetic fields for which the double-well potential exists. Defining the
normalized bias current as 
$\gambbar=\gamb/\gamcn{2}(h)$ with $\gamcn{2}(h)$ given by~\Eq{IcH2simple} 
enables the solutions to~\Eq{cubic} to be written as
\begin{equation}
\label{eq:xsol}
\tanh x_i = 2 \sqrt{\frac{-\tau_2}{3}} \cos 
\left( \frac{\arccos \gambbar}{3} + \frac{2 n \pi}{3} \right)\,.
\end{equation} The constant~$n$ takes on values from the set~$\{0,1,2\}$ 
for a double-well potential where the microshort is centered at~$x=0$.
The set elements correspond respectively to the right potential
minimum~$\ket{R}$ at~$x_3$, the left minimum~$\ket{L}$ at~$x_1$, 
and the potential maximum at~$x_2$.

An analytical expression for the attempt frequency is found in the case of
zero bias current; Substituting~\Eq{xsol} with $\gambbar=0$ and $n=1$ 
into~\Eq{w} leads to
\begin{equation}
\label{eq:wanalytic}
\w = \frac{\sqrt{\mu \bar{h} \left(1-\bar{h}\right) \left(33-\bar{h}\right)}}
{2\left(3+\bar{h}\right)}\,.
\end{equation}
For large 
fields~$\bar{h}\lesssim 1$, near the field at which the microshort-induced
potential barrier disappears, a series expansion of the square 
of~\Eq{wanalytic} about~$\bar{h}=1$ reveals that the attempt frequency
behaves as
\begin{equation}
\label{eq:wh}
\w \simeq \sqrt{\mu\epsh/2}
\end{equation}
where $\epsh\equiv 1-\bar{h}$.

The effect of bias currents just below the critical 
current, $\gambbar \lesssim 1$, on the attempt frequency is of interest when
interpreting microwave spectroscopy data for a vortex in a metastable
state. Such data provide information on the shape of the potential well.
For large magnetic fields, $\bar{h} \lesssim 1$, and large bias currents
the curvature of the potential minimum can be described by
\begin{equation}
\label{eq:curvature}
U''(x_1) \simeq 4 \mu \epsh \sqrt{2\epsg/3}
\end{equation}
where $\epsg\equiv 1-\gambbar$.
On inserting \Eq{curvature}
into \Eq{w}, one finds that the attempt frequency for a vortex pinned 
by a microshort in a metastable potential well depends on the bias current as
\begin{equation}
\w \simeq \sqrt{\mu\epsh}\,(\epsg /6)^{1/4}\,.
\end{equation}
Just as for a vortex pinned by a microresistor in the absence of magnetic
field,\cite{kato96,wallraff00phd} the small oscillation frequency~$\w$ varies
with bias~$\gambbar$ as the fourth root of the term $1-\gambbar$.

\subsection{Quantum properties}

To investigate the quantum properties of an annular Josephson junction with 
a lithographic microshort, we start with the normalized Euclidean action~$S$, 
in units of~${\cal E}_0/\omegaP$, at zero temperature:
\begin{multline}
S[\varphi(x,\tau)]=\int_0^\infty d\tau \int_{-l/2}^{l/2} dx \\
\left\{ 
\frac{\varphi_\tau^2}{2}+ 
\left[1+\frac{\mu}{4} \delta(x) \right] \left(\frac{\varphi_x^2}{2}+
1-\cos{\varphi} \right) + \gambtotal \varphi 
\right\}\,.
\label{eq:actE}
\end{multline}
Here the time variable is transformed as $t=i\tau$, and
the vortex coordinate $x_0(\tau)$ is time dependent:
\begin{equation} 
\varphi_v(x,\tau)=4 \arctan\{\exp[x-x_0(\tau)]\}\,.
\end{equation}
After integration of~\Eq{actE} over the spatial coordinate~$x$,
the effective action has the form
\begin{equation} 
S[\varphi(x,\tau)]=\int_0^\infty d\tau 
\left[  \frac{\vortexmass \dot{x}_0^2}{2} + U(x_0) \right] 
\end{equation}
where $U(x_0)$ is given by~\Eq{Ux0}.

At zero bias current, the tunnel splitting~$\hbar\Delta_0$ 
depends on the instanton action~$S$ as\cite{weiss99}
\begin{equation}
\label{eq:coupling}
\hbar\Delta_0 = 4 \w \hbar \sqrt{\frac{3S}{2\pi\hbar}} 
   \exp\left(-\frac{S}{\hbar}\right)
\end{equation}
where
\begin{equation} 
\label{eq:action}
S=\frac{16 \Delta U}{3 \w} 
\end{equation}
and $\Delta U$ denotes the height of the
potential barrier. From~\Eq{Ux0} and~\Eq{xsol} with $n=1$,
\begin{align}
\Delta U &= \mu(\tanh^2 x_1-\bar{h} x_1^2) \\
&\simeq 3 \mu \epsh^2 /8 \quad \text{for} \quad \bar{h} \lesssim 1\,.
\end{align}
Hence the action varies
with large magnetic fields~$\bar{h}$ as
\begin{equation}
S=2\sqrt{2\mu} \epsh^{3/2}\,,
\end{equation}
and the quantum tunneling rate~$\Delta_0$, in units of $\omegaP$, is given by
\begin{equation}
\Delta_0=2\sqrt{\frac{3}{\pi\hbar}} \left(2\mu\right)^{3/4} \epsh^{5/4} 
\exp\left(-\frac{2\sqrt{2\mu} \epsh^{3/2}}{\hbar}\right)\,.
\label{eq:qrate}
\end{equation}

\subsection{Operation as a two-level system}
\label{sec:tls}

Coherent oscillation of the vortex between the basis states~$\ket{L}$
and~$\ket{R}$ is controlled by short pulses of the in-plane
field~$H$. At zero bias current, these states are degenerate. Oscillation
takes place during a positive pulse, when the 
potential barrier $\Delta U$ is lower and the interwell coupling~$\Delta_0$ is
thereby increased. At this field, the potential minima are close enough to 
each other that the crossover temperature~$T_0$, 
below which exponential relaxation gives way to underdamped coherent
oscillation, is greater than the temperature~$T$ of the qubit.\cite{leggett87}
After the end of the pulse, the crossover temperature~$T_0$ is lower than the 
qubit temperature~$T$, and the oscillation has ceased. Readout should take
place before overdamped relaxation causes the vortex to transition to the
other well. At this point, there is enough time to further lower the field 
if it is necessary to reduce the barrier height before the depinning current of 
the $\ket{L}$~state can be measured.

We now consider the feasibility of experiments in 
the quantum regime using realistic trilayer fabrication parameters.
We assume a critical current density of $\jc=\unit[1]{kA/cm^2}$, 
a junction width of $w_0=\unit[0.6]{\mu m}$, and an
additional junction area of $(w_\mu-w_0) \lmu=\unit[0.2]{\mu m^2}$.
This corresponds to a Josephson length of $\lambdaJ\approx\unit[12]{\mu m}$,
a microshort strength of $\mu = 0.1$, and a reduced
Planck's constant of $\hbar =\unit[3.5 \times 10^{-3}]{{\cal E}_0/ \omegaP}$.
In normalized units, the magnitude of the reduced Planck's constant~$\hbar$
is inversely proportional to the junction width~$w_0$. Hence quantum effects
are more readily measured in narrower junctions.~\cite{abdumalikov06}
The length of the junction is taken to be $\unit[30]{\lambda_J}$.

\begin{figure}
\resizebox{\figwidth}{!}{\includegraphics{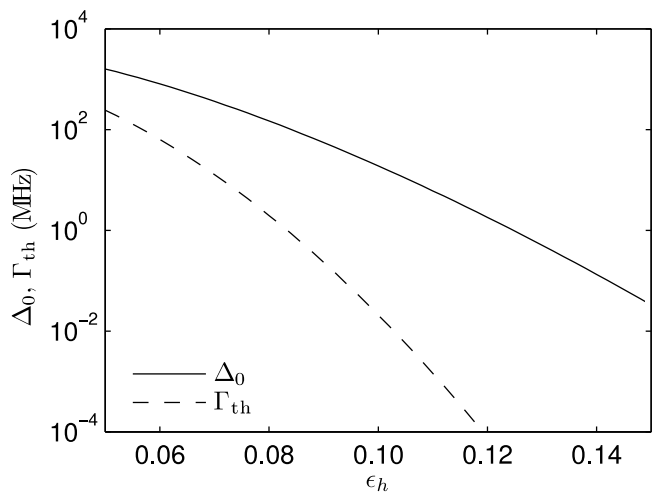}}
\caption{Quantum tunneling rate~$\Delta_0$ (solid line) and upper limit 
for the thermal activation rate~$\thrate$
(dashed line) for a junction of critical current density 
$\jc=\unit[1]{kA/cm^2}$, width~$w_0=\unit[0.6]{\mu m}$, and microshort
strength $\mu=0.1$. \label{fig:rate}}
\end{figure}

We estimate an upper limit for the
thermal escape rate~$\thrate$ by assuming moderate damping:
\begin{equation}
\label{eq:thesc}
\thrate=\frac{\w}{2\pi}\exp\left(-\frac{\Delta U}{\kb T}\right)\,.
\end{equation}
The actual thermal activation
rate is expected to be substantially slower because the junction will be in 
the low damping regime. \Figure{rate} aids comparison of the upper limit for
the thermal escape rate~$\thrate$ at temperature $T=\unit[50]{mK}$ with 
the quantum tunneling rate~$\Delta_0$ 
given by~\Eq{qrate}. Clearly, thermal
activation should be insignificant during operation of the junction
as a two-level system at $T=\unit[50]{mK}$.

The friction coefficient~$\eta$ plays a crucial role in 
the dissipative dynamics of a two-level system. It is given by
\begin{equation}
\eta=\frac{1}{\omega_P R C} 
\end{equation}
where $C$ denotes the junction capacitance and $R=\unit[50]{\Omega}$,
the real part of the impedance of the microstrip biasing lines
feeding the junction. For our chosen junction parameters, the 
classical friction coefficient is
$\eta = 2\times 10^{-3}$.

\begin{figure}
\resizebox{\figwidth}{!}{\includegraphics{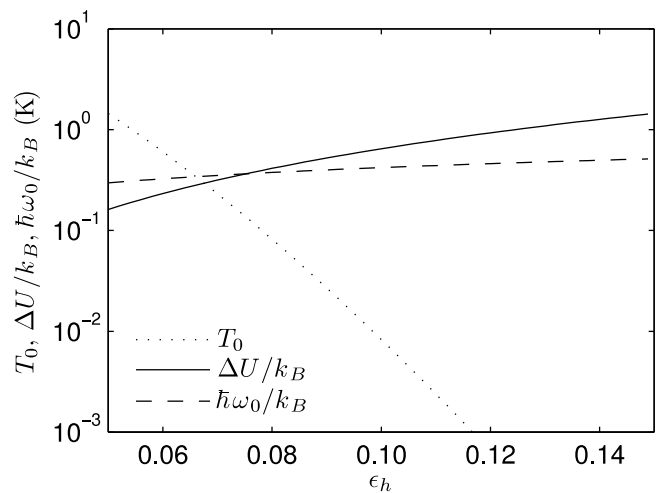}}
\caption{Barrier height~$\Delta U$ (solid line); 
spacing~$\hbar \w$ (dashed line) between the first excited state and the 
ground state in each well; and
  crossover temperature~$T_0$ (dotted line), below which exponential relaxation
  becomes underdamped coherent oscillation. The 
  junction parameters are the same as for~\Fig{rate}.
  \label{fig:temperatures}}
\end{figure}

To determine the experimental conditions under which coherent oscillation
of the vortex between the left~$\ket{L}$ and right~$\ket{R}$ potential minima 
is observable, we use the  
dimensionless system-environment coupling strength
\begin{equation}
\alpha=\frac{\eta (\Delta x)^2}{2\pi \hbar}
\end{equation}
where $\Delta x$ is the distance between the minima according to~\Eq{xsol}. 
The crossover temperature~$T_0$ depends on the coupling strength~$\alpha$ 
as\cite{leggett87}
\begin{equation}
T_0=\frac{\hbar}{\kb} \frac{\Delta_0}{\alpha \pi}\,.
\end{equation}
\Figure{temperatures} displays the crossover temperature~$T_0$,
the barrier height~$\Delta U$, and the spacing~$\hbar \w$ between the first 
excited state and the ground state in each well.
At a temperature of $T=\unit[50]{mK}$, an applied field of
$\epsh\sim 0.06$ for the quantum operation and
$\epsh\sim 0.14$ for the readout should yield a measurable quantum tunneling 
rate of $\Delta_0 \sim \unit[700]{MHz}$ and allow a readout time of up to
tens of milliseconds. Rectangular field pulses of the required height 
$\Delta \bar{h}=0.08$ and frequency could be produced using a microstrip 
or the near field of an rf antenna. During quantum operation at 
$\epsh\sim 0.06$, the level spacing~$\hbar\w$ is one order of 
magnitude greater than the tunnel splitting~$\hbar \Delta_0$. Thus pulse
frequencies comparable to the tunneling rate $\Delta_0$ can be used without
excitation of
the vortex to a higher level provided that the rise time of the field pulse
is longer than~$2 \pi/\w$.



\section{Experiment}
\label{sec:results}

\subsection{Samples and setup}
\label{sec:setup}

The results presented here pertain to two nominally identical long Josephson 
junctions of the geometry depicted in~\Fig{lith}. The junctions were 
fabricated on different chips within the same run 
of a standard Nb-AlO$_x$-Nb trilayer process\cite{hypres} and had a 
critical current density of~$\jc = \unit[1.2]{kA/cm^2}$ at temperature 
$T = \unit[1.4]{K}$, which corresponds to a Josephson length of 
$\lambdaJ=\unit[13]{\mu m}$ (after including the effect of the idle 
region).\cite{franz01} The basic shape of the junctions was annular, with a
radius of~$r=\unit[50]{\mu m}$ and a width of~$w_0=\unit[2.4]{\mu m}$. A short 
length $\lmu=\unit[1.4]{\mu m}$ of each junction was wider, 
$w_{\mu}=\unit[2.9]{\mu m}$, forming a lithographic microshort of expected 
strength~$\mu=0.09$. The junctions had Lyngby style\cite{davidson85} electrode 
leads, whose width slightly exceeds the junction diameter~$2r$. 

Each sample was situated within a solenoid which produced the in-plane magnetic 
field~$H$. For measurements of the vortex depinning current~$I_C(H)$ at 
$T=\unit[1]{K}$, the solenoid was attached to the cold finger of a dilution 
refrigerator. The junction was biased through low pass RC filters at
the $\unit[1]{K}$ stage.\cite{wallraff03a} Measurements at $T=\unit[1.4]{K}$ 
were carried out in a $^4$He cryostat. For both setups,
room temperature feedthrough capacitor filters were used. The Josephson vortex 
was trapped upon cooling through the critical temperature of the niobium 
electrodes. Its depinning current was evaluated as the product of the known
bias current ramp rate with the time elapsed between the ramp passing
through zero current and the appearance of voltage across the junction 
electrodes.\cite{wallraff03a}

\subsection{Bistable states}

\begin{figure}[htbp]
\resizebox{\figwidth}{!}{\includegraphics{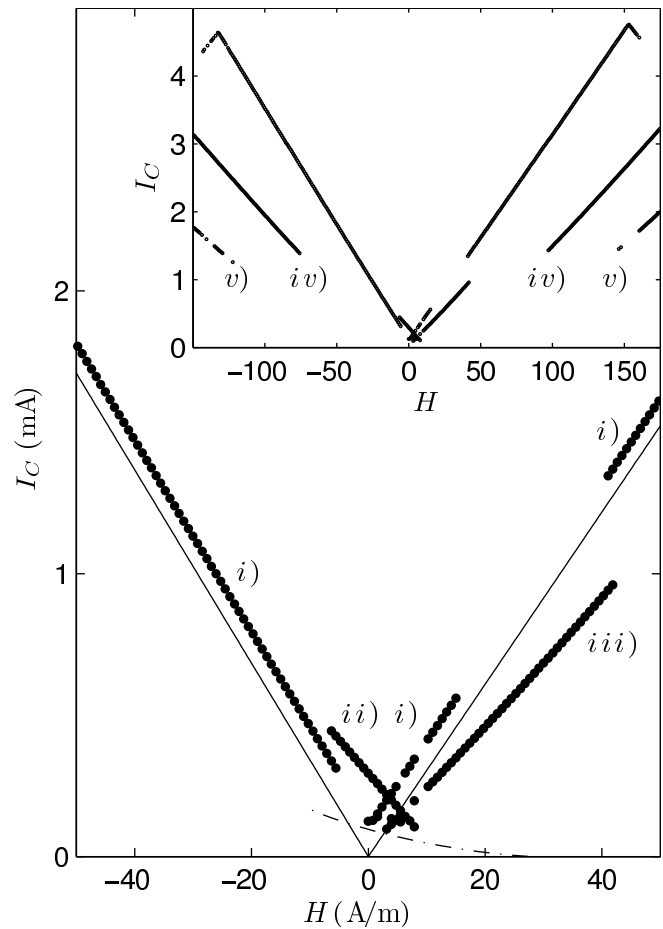}}
\caption{Vortex depinning current~$I_C$ for \jb as a function of applied
magnetic field strength~$H$ at temperature~$T=\unit[1.4]{K}$. The standard 
deviation, which is on the order of $\unit[1]{\mu A}$, is not shown. 
Sections of the data (black circles) form approximately straight lines which
originate in the escape of the vortex over a potential barrier induced by 
either $\romannumeral1)$ the magnetic field, $\romannumeral2)$ the 
microshort, or $\romannumeral3)$ the injectors.\cite{ustinov02, malomed04}
The depinning currents expected from~\Eqs{IcH1}{IcH2} for the magnetic field
and microshort barriers are also plotted (solid and dot-dashed lines, 
respectively). In calculating the depinning current from the magnetic field
barrier, the self-field of the bias current was 
taken into account. The inset displays the supercurrent maxima of the 
$I_C(H)$~pattern as well as branches due to the nucleation of
$\romannumeral4)$ one or $\romannumeral5)$ two vortex-antivortex pairs.
}
\label{fig:exp-theory}
\end{figure}

Measured data on the dependence of the vortex depinning current~$I_C$ as a 
function of the external magnetic field strength~$H$ are plotted in 
\Fig{exp-theory} together with \Eqs{IcH1}{IcH2}. The $I_C(H)$~data consists of 
lines 
which are labelled according to the source of the potential barrier over which 
the vortex escaped: $\romannumeral1)$ the magnetic field, $\romannumeral2)$ 
the lithographic microshort, and $\romannumeral3)$ the 
injectors.\cite{ustinov02, malomed04} The branches which appear at high 
field~$H$ originate from the nucleation of $\romannumeral4)$ one or 
$\romannumeral5)$ two vortex-antivortex pairs.\cite{martucciello96b} 
The injectors are leads for local current injection, which provide a means 
of inserting a vortex into the junction on demand. They were not used in the 
experimental work described in this article and they do not affect the shape 
of the relevant part of the double-well potential, even though the injector 
branch is present at weak positive fields~$H$. This is because the current 
injection leads are located at $x=\pi r$, as far from the microshort as 
possible.

The dependence of the depinning current on the magnetic-field 
barrier is typical for long angular junctions. At zero field the barrier
is absent and the junction critical current, at which the vortex depins,
is therefore minimal. Increasing the magnetic field strength leads to a larger
vortex depinning current as expected from~\Eq{IcH1}, which is plotted as a 
solid line in \Fig{exp-theory}. In order to graph \Eq{IcH1}, the geometrical
parameter~$\kappa$ was calculated from the gradient of the branches
labeled~$\romannumeral1$ while taking into account the self-field of the 
bias current. The maximum critical current was measured independently, and 
the values of the other junction parameters came from the specifications of 
the fabrication process.

Linear extrapolations of branch~$\romannumeral1$ for positive and
negative field~$H$ do not intersect at zero critical current. This could be due 
to the self-field of a current induced by a small magnetic field~$\htrans$
oriented perpendicular to the junction plane. The induced current 
circulates along the electrode rings. At $x_0= +\pi r/2$ and $x_0= -\pi r/2$,
where the vortex depins from the barrier induced by positive and negative 
field~$H$, respectively, the self-field of the circulating current is of
opposite polarity. The contribution of this self-field to the total in-plane 
field~$H$ shifts the measured branches~$\romannumeral1$, for both positive 
and negative field~$H$, along the $H$ axis towards the origin.

For negative field polarity, the magnetic-field barrier is centered at the 
microshort, at $x_0=0$, which results in a single-well potential. For 
positive polarity, the magnetic-field barrier, at $x_0=\pi r$, is located 
diametrically opposite the microshort barrier; Together they form the 
double-well potential. As seen in \Fig{exp-theory}, at temperature 
$T=\unit[1.4]{K}$ the stable vortex states of the double-well potential are
distinguishable by means of the depinning current over the field
interval $0 \lesssim H \lesssim \unit[5]{A/m}$. Depinning currents from both 
of the two branches $\romannumeral1$ and $\romannumeral2$ are also measured for 
fields near $H=\unit[-4]{A/m}$ although the potential at zero bias
consists of a single well. In this case the junction radius was large
enough that, at finite bias, a second local potential minimum formed
between the positions where the respective gradients of the field and 
microshort contributions to the potential, $\partial{U^h}/\partial{x}$ and 
$\partial{U^\mu}/\partial{x}$, were maximal.

In \Fig{exp-theory}, the measured 
depinning current from the microshort barrier is several times larger
than predicted by \Eq{IcH2}, which is represented 
by a dot-dashed line. We discuss this issue in \Sec{dis2},
where we show how an
external magnetic field~$\htrans$ perpendicular to the junction plane can
result in an effective microshort strength~$\mueff$ which grows with
increasing field strength~$\htrans$. When $\mueff \not \approx \mu$, the
dependence of the vortex small oscillation frequency~$\w$ on the in-plane 
field~$h$ differs from \Eqs{wanalytic}{wh}. This prevents direct comparison
of the data in \Figs{temperatures}{exp-theory}. One means of avoiding
interaction between the microshort and transverse field~$\htrans$ is 
suggested in \Sec{improved_design}.

\subsection{State preparation}

\begin{figure}[htbp]
\resizebox{\figwidth}{!}{\includegraphics{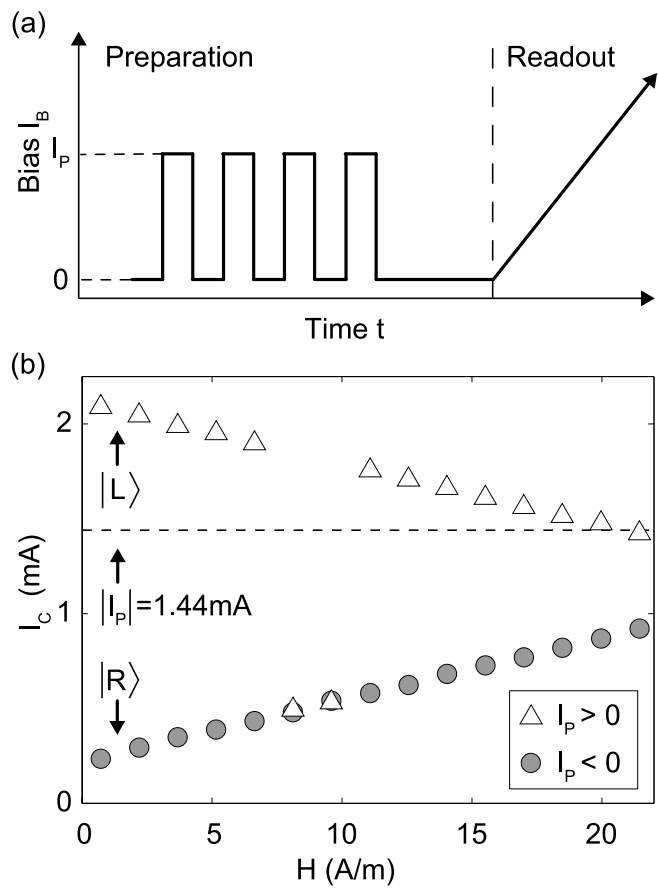}}
\caption{State preparation by means of a series of shaker\cite{lefevre92} 
bias current pulses applied to \ja at 
temperature~$T=\unit[1]{K}$.\cite{kemp06phd}
(a) The bias current during state preparation and readout.
(b) Positive (white triangles) and negative (grey circles) preparation pulses
place the vortex in the left~$\ket{L}$ and right~$\ket{R}$ potential wells,
respectively. The magnitude
of the preparation pulses (dashed line) is indicated.}
\label{fig:shaker}
\end{figure}

To base a quantum bit on this junction design, it is necessary to be
able to produce a defined initial state. We prepared the vortex as 
desired in the
left~$\ket{L}$ or right~$\ket{R}$ potential well by means of a
shaker\cite{lefevre92} bias current sequence,\cite{kemp06phd} which is 
depicted in~\Fig{shaker}a. 
The amplitude~$|I_P|$ of the preparation pulses is such that a pulse
frees the vortex only when it is pinned by the magnetic field and
not when it is pinned by the microshort. The vortex remains
pinned in the left~$\ket{L}$ and right~$\ket{R}$ wells for pulses of positive 
and negative polarity, respectively. 
With each successive pulse, the likelihood of locating the vortex in the 
chosen well increases. The number of pulses needed
to reliably attain the desired state depends on the probability of
the vortex being retrapped in the chosen well. Note that the vortex can also
be prepared deterministically in the left~$\ket{L}$ or right~$\ket{R}$ well
by applying a bias pulse of the appropriate polarity at zero field
and increasing the external field afterwards, but this procedure is 
slower due to solenoid inductance. 

After the preparation stage, the bias current is increased
at constant positive rate in order to measure the vortex depinning
current and identify which potential well the resting vortex was
located in. The readout results\cite{kemp06phd} for thirty preparation pulses 
of amplitude~$|I_P|=\unit[1.44]{mA}$ in \ja at temperature~$T=\unit[1]{K}$ 
are plotted in~\Fig{shaker}b. The white triangles denote depinning currents
measured after preparing the state with positive polarity pulses,
and the grey circles, negative polarity pulses. 

We observed over
tens of trials that thirty preparation pulses are enough to prepare 
a given state at nearly all external magnetic field strengths. An exception
occurred near~$\unit[10]{A/m}$ where the vortex could not be prepared in the
left well~$\ket{L}$ due to the low probability for it to be retrapped 
there. We found that the probability for the vortex to be retrapped in a
particular well is temperature and field dependent. The temperature 
dependence of the retrapping probability may be caused by the damping
changing with temperature. Initialization at a 
field~$H$ where the probability of retrapping the vortex in the desired well
is high offers the advantage that fewer preparation pulses are required. 
We expect that the shaker initialization process can be used at most fields
where the depinning currents from the~$\ket{L}$ or~$\ket{R}$ wells are
distinct as long as the amplitude~$|I_P|$ of the preparation pulses
lies between these currents and is greater than the depinning current
associated with any other potential barriers in the junction. Therefore
\jb could be initialized with this procedure over a field range from
$H=\unit[0]{A/m}$ to $H\sim\unit[5]{A/m}$, a shorter interval than for \ja.

In \Fig{shaker}b, the measured depinning current from the left well~$\ket{L}$, 
where the vortex is pinned by the microshort, is an order of magnitude 
greater than predicted by~\Eq{IcH2}. Such a large enhancement occurs when 
magnetic flux is trapped around both electrodes. The flux-induced supercurrent
interacts with the lithographic microshort as described in the next section.

\section{Discussion}
\label{sec:disc}

\subsection{Enhanced vortex pinning due to flux trapped around both electrodes}
\label{sec:dis}

We have found that the vortex depinning current strongly
depends on the strength of a magnetic field~$\htrans$ applied perpendicular to
the junction plane during cooling below the superconducting transition 
temperature $T_C\sim \unit[9.2]{K}$. After field cooling the transverse 
field~$\htrans$ was switched off, and then the critical current~$I_C$ was 
measured as a function of the in-plane field~$H$. 
We repeated this process for a range of transverse fields~$\htrans$ and obtained
several distinct $I_C(H)$~patterns. Each $I_C(H)$~pattern was observed for a
particular subinterval of transverse field. Reversing the polarity
of the applied transverse field reflected the $I_C(H)$~patterns about
the $I_C$~and $H$ axes simultaneously. Further results
pertaining to field cooling annular junctions under perpendicular fields
are presented elsewhere.\cite{price08,monaco08}

\begin{figure}[htbp]
\resizebox{\figwidth}{!}{\includegraphics{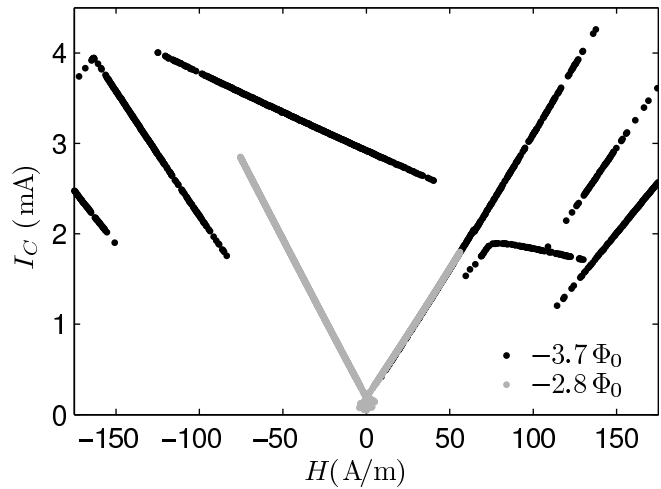}}
\caption{Vortex depinning current~$I_C$ versus in-plane magnetic field~$H$
at $T=\unit[1.4]{K}$, obtained after field cooling \jb under transverse 
magnetic fields equivalent to $-2.8$ (grey) and $-3.7$ (black) flux quanta 
through each electrode loop. The perpendicular field was switched off before 
the $I_C(H)$~pattern was recorded.
}
\label{fig:cycling}
\end{figure}

\Figure{cycling} shows $I_C(H)$~patterns which were obtained after
field cooling under transverse fields~$\htrans$ differing by 
around~$\unit[0.2]{A/m}$ and 
equivalent to fluxes of $\Phi_Z=\unit[-2.8]{\PhiZ}$ (grey) and 
$\Phi_Z=\unit[-3.7]{\PhiZ}$ (black) through each electrode loop. 
Whereas the $I_C(H)$~pattern obtained
after field cooling under~$\Phi_Z=\unit[-2.8]{\Phi_0}$ is typical of a single
trapped vortex, the pattern for $\Phi_Z=\unit[-3.7]{\Phi_0}$ 
resembles~\Fig{shaker}b in that the vortex depinning current over 
the microshort-induced barrier is enhanced by an order of magnitude.
The difference between these two $I_C(H)$~patterns originates in 
the amount of flux trapped within both electrode rings.
Note that the microshort depinning current measured after field cooling
under $\Phi_Z=\unit[-2.8]{\PhiZ}$ does not agree with \Eq{IcH2}, possibly 
due to the influence of a weak background transverse field.

Field transverse to the junction plane
induces a supercurrent~$I_\Phi$ which circulates around the 
electrode loop. The fraction of supercurrent in each electrode is in the 
inverse ratio of their inductance per unit length. Due to the fabrication 
process, this ratio changes at the microshort boundaries, where not only
the active region but also the neighboring part of the upper electrode widens
as illustrated in \Fig{qubit}. Consequently, some supercurrent~$\varepsilon$ 
passes upwards through the junction barrier on one side of the microshort and 
downwards on the other. The corresponding spatial dependence of the
Josephson phase $\vphi(x)$ is equivalent to that of a current 
dipole\cite{ustinov02, malomed04} or fractional 
vortex,\cite{goldobin04, goldobin04a} objects which can increase the vortex 
depinning current over the microshort-induced potential barrier. For a 
pointlike ($\lmu \ll \lambdaJ$) dipole of strength 
$\kappa_\mu=\varepsilon\lmu$, \Eq{pert} becomes\cite{malomed04}
\begin{equation}
\gambtotal(x)= \gamb + \frac{\partial{h_R}}{\partial{x}} 
   + \kappa_\mu \delta'(x) \, .
\end{equation}
The contribution of the lithographic microshort to the vortex potential is then
\begin{align}
\U^\mu(x_0)&=\mu \sech^2 x_0 - 2\kappa_\mu \sech x_0 \\
    &= (\mu - 2\kappa_\mu) \sech^2 x_0 - \kappa_\mu x_0^2 + O(x_0^4) \, .
       \label{eq:Umicrod}
\end{align}
A negative dipole strength~$\kappa_\mu$, which corresponds to the induced
supercurrent~$\varepsilon$ circulating in the direction of decreasing~$x$, 
thus increases the height of the potential barrier at the lithographic 
microshort.

The order-of-magnitude enhancement of the microshort depinning current evident 
in~\Fig{shaker}b is due to the dipole current~$\varepsilon$ induced by
magnetic field perpendicular to the junction plane. The field is from
flux quanta trapped within the electrode rings, possibly in combination with 
a residual background
field~$\htransZ$. The background field for the data presented in 
\Figs{exp-theory}{shaker} is likely to be different because they were 
recorded in different cryostats. An increase in the depinning current
due to trapped flux takes on discrete values, whereas the increase
caused by a background field does not and can therefore be much smaller.
Another difference in the effect of trapped flux and a background field 
is that they induce dipole currents~$\varepsilon$ of opposite polarity.

\subsection{Enhanced vortex pinning due to magnetic field applied
transverse to the junction plane}
\label{sec:dis2}

\begin{figure}[htbp]
\resizebox{\figwidth}{!}{\includegraphics{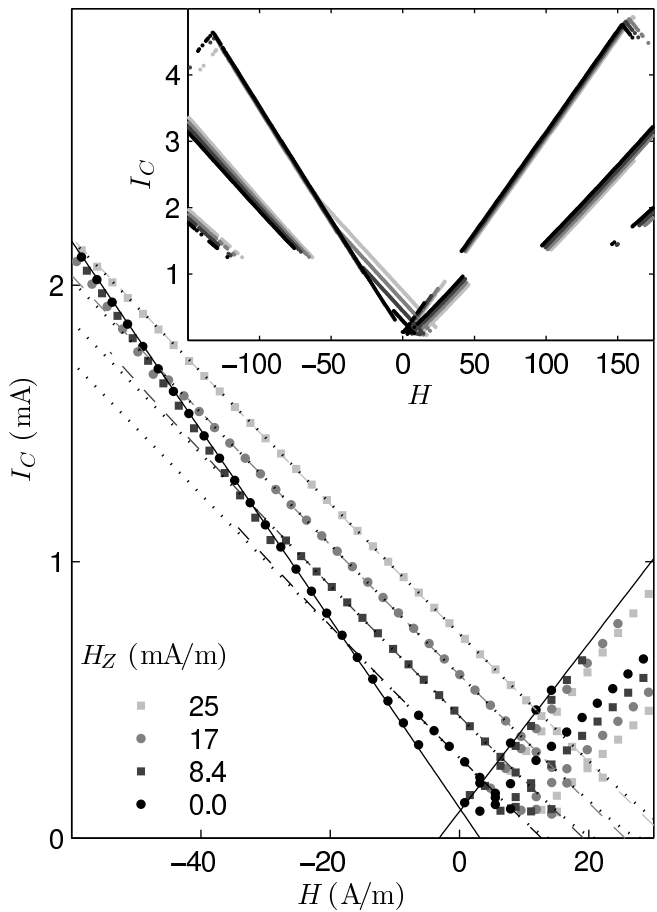}}
\caption{Vortex depinning current~$I_C$ as a function of applied in-plane
magnetic field~$H$ and perpendicular field~$\htrans$. Measurements were 
recorded at a temperature of~$T=\unit[1.4]{K}$ using~\jb.  The microshort
depinning current at $H=\unit[0]{A/m}$ depends linearly on the applied 
perpendicular field~$\htrans$.  Also plotted is a linear
fit (solid line) to the magnetic field depinning current for 
$\htrans=\unit[0]{A/m}$. The fitted depinning
current from the microshort barrier is indicated for two cases of 
misalignment of the solenoid field~$H$: $\thetayz$ out of 
(dotted line) and~$\thetaxy$ within (dashed line) the junction plane. 
The inset displays the $I_C(H)$~pattern as a function
of~$\htrans$ over a field~$H$ interval which includes the supercurrent
maxima.
}
\label{fig:fit-data}
\end{figure}

We have investigated the effect on the microshort depinning current of
a uniform external field~$\htrans$ applied transverse to the plane of a 
junction in the superconducting state. The transverse 
field~$\htrans$ was generated using a current coil located underneath the chip.
Note that, since the field lines produced by the coil are screened by
the superconducting electrodes and cannot pass through the hole of the 
electrode loop, deflected lines concentrate near the outer
edges of the electrodes.\cite{monaco07,monaco08a} Thus the magnitude of the
local external field tangential to the electrode surface is substantially
larger than~$\htrans$. In the analogous situation of a disc of radius~$r$ and
thickness~$a$ in the Meissner state in a uniform transverse field, the
field at the edge of the disc is a factor 
of~$(1+\sqrt{2r/a})$ greater.\cite{benkraouda96}

\Figure{fit-data} displays the 
$I_C(H)$ patterns recorded for various perpendicular field strengths~$\htrans$
applied to the same initial single-vortex junction state, which was obtained
by cooling in zero field. An offset in the in-plane field~$H$, due to
deflection of the applied transverse field~$\htrans$ 
through the tunnel barrier, has been removed from each $I_C(H)$ pattern.
The approximately straight lines of the patterns
are associated with the vortex escaping over a potential barrier due
to either the magnetic field~$H$, the microshort, or the injectors as 
indicated in~\Fig{exp-theory} for zero perpendicular field~$\htrans$. The 
measured microshort depinning current clearly increases with increasing
magnitude of the circulating supercurrent induced by the perpendicular
field.

The dipole strength~$\kappa_\mu$ depends on the applied transverse 
field~$\htrans$ as $\kappa_\mu=-\zeta(\htrans+\htransZ)$ where~$\zeta$ is 
a proportionality constant. For small~$\kappa_\mu$ or~$x_0$, 
\Eq{Umicrod} approximates to $\U^\mu(x_0)=\mueff \sech^2 x_0$
where~$\mueff$ is an effective microshort strength that varies linearly 
with the transverse field~$\htrans$:
\begin{equation}
\mueff=\mu + 2 \zeta (\htrans+\htransZ) \, .
\end{equation}
With this assumption,
we fitted the microshort depinning current to~\Eq{IcH2} and determined
the magnitude of the background perpendicular field~$\htransZ=\htransZV$ and
the constant~$\zeta=\zetaV$ using the method
of least squares. An additional fit parameter was the degree of 
misalignment of the solenoid field~$H$.

The measured microshort depinning current depends more strongly on the
in-plane field~$H$ than expected from~\Eq{IcH2}, as is seen from 
\Fig{exp-theory}. We believe that this was caused by a 
misalignment of the solenoid field~$H$. As is shown in \Fig{fit-data}, 
a misalignment of either~$\theta_1=\thetayz$ out of 
(dotted lines) or $\theta_2=\thetaxy$ within (dashed lines) the junction plane 
fits the observed $H$~dependence of the microshort depinning current. 
In both cases, one fit curve is plotted for each of the four values
of perpendicular field~$\htrans$ under which $I_C(H)$~patterns were recorded.
That the gradient of the measured microshort $I_C(H)$~curve depends so
sensitively on the degree of misalignment out of 
the junction plane results from the solenoid field~$H$ having a $z$~component
when directed out of the plane. The effective microshort strength~$\mueff$
is then a function of this component $H\sin\theta_1$ as well as the
intentionally applied perpendicular field~$\htrans$. The larger than 
predicted gradient
of the microshort $I_C(H)$~curve could also be due to a simultaneous
misalignment of the solenoid field~$H$ both within and out of the 
junction plane. Hence the discrepancy seen in~\Fig{exp-theory}
between the measured microshort depinning current and the prediction
of the one-dimensional model, given by~\Eq{IcH2}, can be attributed to
a misalignment of the solenoid field~$H$ in addition to the presence of a small
background perpendicular field $\htransZ=\htransZV$. One means of avoiding
interaction between the lithographic microshort and transverse 
field is described in the next section.

\subsection{Decoherence sources and improved design}
\label{sec:improved_design}

The vortex qubit has been predicted to have a long coherence time for a 
superconducting qubit, on the order of tens of microseconds.\cite{kim06}
This estimate considered the effect of quasiparticle dissipation and
weakly fluctuating critical and bias currents in a long linear junction. Low
frequency critical current noise was found to lowest order not to alter the
shape of the vortex potential and hence not to contribute to decoherence.
Since the equilibrium density of quasiparticles is exponentially small at 
low temperatures, the coherence time was predicted to be limited by
low-frequency bias noise.

Additional factors which could reduce the coherence time in our 
proposed qubit are increased sensitivity to flux noise, the presence of
excess quasiparticles, and interaction with two-level systems 
in the dielectric. Flux fluctuations in the $y$ or $z$ directions modulate the 
barrier height of the microshort qubit, causing exponential change in the
tunneling frequency. Therefore it is important to minimize flux noise. Excess
quasiparticles are generated when the junction switches to the voltage state.
Their contribution to shot noise in the bias current can be decreased with
quasiparticle traps.\cite{lang03} Two-level systems are thought to be 
a major source of decoherence in superconducting qubits.\cite{martinis05} 
They originate from charge fluctuations and couple to the qubit via the 
electric 
field~$\dot{\varphi}/t_j$\cite{martin05} where $t_j$ is the tunnel barrier 
thickness. In a microshort qubit, two-level systems near the microshort
may couple to the qubit via the phase oscillation which results
when the vortex oscillates between the $\ket{L}$ and $\ket{R}$ states. The
amplitude of the phase oscillation is largest at the microshort and decays
over a few Josephson lengths. The number of two-level systems which
interact with the microshort qubit could be reduced by employing a
tunnel barrier of $a$-Si:H,\cite{oconnell08} SiN$_x$,\cite{martinis05} or 
epitaxial Al$_2$O$_3$.\cite{oh06}

Decoherence due to flux noise transverse to the junction plane is 
avoided in a microshort qubit without an electrode loop. For example,
an improved qubit design is a long junction which consists of two linear
segments separated by an annular segment that contains the lithographic 
microshort. At the place where the junction centerline transitions from 
annular to straight, the field-induced potential 
barrier levels off as seen from \Eq{fconv} because, in the straight
segments, the component $h_R$ of the in-plane field transverse to the
junction centerline is constant. The location of the transition 
is chosen so that, at the readout field, the
microshort depinning current~$\gamcn{2}$ is greater than the field
depinning current~$\gamcn{1}$. This enables state detection with an
RS flip flop at one end of the junction\cite{kaplunenko04} provided that the 
readout bias permits the vortex to overcome the field-induced 
potential barrier but not the microshort-induced barrier. At the other
end of the junction is a single flux quantum (SFQ) generator for qubit 
initialization.\cite{kaplunenko04} As well as insensitivity to transverse
field noise, this design offers the advantage of faster readout, during
which fewer excess quasiparticles are generated. Also, the length of
the junction no longer constrains the radius of the annular part; By 
choosing a shorter radius~$r$, control pulses of a given 
amplitude~$\Delta \bar{h}$ are achieved with smaller amplitude pulses of
the in-plane field~$H$. Using in-plane field pulses large enough that
the potential transforms from a double to a single well could reduce
the sensitivity of the coherent oscillation frequency to flux
fluctuations in the $y$~direction.\cite{poletto09}

\section{Conclusion}

We have studied a vortex qubit based on an annular Josephson junction
containing a lithographic microshort. From the one-dimensional vortex
potential, we have derived the magnetic field dependence of the vortex
depinning current over a microshort-induced potential barrier. We have also
obtained the frequency of vortex oscillation within the well. From this 
we find that, for a low microshort-induced potential barrier, the 
exponent in the coupling~$\Delta_0$
between degenerate minima varies with field as~$(1-\bar{h})^{3/2}$. 

The proposed vortex qubit design has been tested experimentally in the
classical regime.
We observed bistable vortex states located on either side of the microshort.
Preparation of the vortex in a given potential well was
achieved by means of a shaker sequence of bias current pulses.
We noticed that the depinning current from a lithographic
microshort can be enhanced by flux trapped around both superconducting 
electrodes as well as field applied perpendicular to the junction plane.

\begin{acknowledgments}
We thank Edward Goldobin for valuable discussions. This work was partially 
supported by the `European Superconducting Quantum Information Processor'
(EuroSQIP) project, the European Science Foundation Research Networking 
Programme `Arrays of Quantum Dots and Josephson Junctions' (AQDJJ), 
the Deutsche Forschungsgemeinschaft, and the EPSRC grant No. EP/E042589/1.
\end{acknowledgments}


\end{document}